\documentclass[12pt]{article}
\usepackage{latexsym}
\usepackage{amssymb}
\newcommand{\dintst}{\int \! \! \! \int d^{4}x \, d^{4}y}

\newcommand{\pslash}{\ensuremath{\not \! p}}
\newcommand{\qslash}{\ensuremath{\not \! q}}
\newcommand{\ppslash}{\ensuremath{\not \! p'}}
\newcommand{\ifour}{\int \! \! \frac{d^4 k}{(2 \pi)^4}}
\newcommand{\difour}{\int \frac{d^4 k_1 d^4k_2}{(2 \pi)^8}}
\newcommand{\dintsoT}{\int_0^{s_0(T)} \! \! \! \! \! \! \!\;ds \! \!
  \int_0^{s'_0(T)}\! \! \! \! \! \! \! \;ds'} 
\newcommand{\abp}{|\vec{p}|}
\newcommand{\dintsof}{\int_0^{s_0} \! \! \int_0^{s'_0} \!ds\,\; ds'}

\begin{document}
\thispagestyle{empty}
\mbox{ }
\rightline{UCT-TP-256/99}\\
\rightline{June 1999}\\
\vspace{2.0cm}
\begin{center}
\begin{Large}
{\bf QCD sum rule determination of the axial-vector  
 coupling of the nucleon at finite temperature\\}
\end{Large}
\vspace{.7cm}
{\bf  C. A. Dominguez $^{(a)}$, M. Loewe $^{(b)}$, C. van Gend
$^{(a)}$ $^{(c)}$} \\
\vspace{.7cm}
$^{(a)}$Institute of Theoretical Physics and Astrophysics\\
University of Cape Town, Rondebosch 7700, South Africa\\
\vspace{.5cm}
$^{(b)}$Facultad de Fisica, Pontificia Universidad Catolica
de Chile\\
Casilla 306, Santiago 22, Chile\\
\vspace{.5cm}
$^{(c)}$National Accelerator Centre, Faure 7131, South Africa\\
\end{center}
\vspace{.5cm}
\begin{abstract}
\noindent
A thermal QCD Finite Energy Sum Rule (FESR) is used to obtain
the temperature dependence of the axial-vector coupling of the nucleon,
$g_{A}(T)$. We find that $g_{A}(T)$ is essentially independent of $T$,
in the very wide range $0 \leq T \leq 0.9 \; T_{c}$, where $T_{c}$ is
the critical temperature. While $g_{A}$ at $T=0$ is $q^{2}$-independent,
it develops a $q^{2}$ dependence at finite temperature. We then obtain
the mean square radius associated with $g_{A}$ and find that it diverges
at $T=T_{c}$, thus signalling quark deconfinement. As a byproduct, we
study the temperature dependence of the Goldberger-Treiman relation.
\end{abstract}
\newpage

The possibility of creating a quark-gluon plasma in
relativistic heavy ion collisions has sparked much interest in
theoretical predictions for the onset of this state \cite{REV1}. In addition
to the search for unambiguous processes signalling the formation
of such a plasma, it is also important to understand the temperature
behaviour of hadronic Green's
functions and their associated parameters, viz. masses, widths, couplings,
etc. The general consensus is that hadronic widths depend strongly on the
temperature; in fact they are expected to diverge at some critical
temperature $T_{c}$, thus signalling quark-gluon deconfinement \cite{GAMMAT}
(hadronic widths are to be understood, in this context, as absorption
coefficients determined by the imaginary
parts of two-point functions). Thermal three-point functions also
provide independent evidence for this phase
transition, as the mean square radii happen to increase with increasing
temperature, becoming infinite at $T=T_{c}$ \cite{3PFT}.\\
A recent investigation of the thermal behaviour of the pion-nucleon
coupling, in the framework of both the linear sigma model and QCD sum
rules \cite{GPINNT}, showed that as the temperature approaches $T_{c}$,
$g_{\pi NN}(T)$ vanishes, while the associated radius diverges. Both
$g_{\pi NN}(T)$ and $\langle r^2_{\pi NN} \rangle(T)$ may thus be
interpreted as signals for the deconfinement phase transition. In this
work we shall determine the temperature behaviour of the axial-vector
coupling constant of the nucleon $g_A \equiv g_A(q^{2}=0)$, and the 
associated radius, using the method of thermal QCD sum rules
\cite{QCDSRT}; specifically,
the leading dimension Finite Energy Sum Rule (FESR).
However, we shall first discuss our own determination of $g_{A}$
at $T=0$, as previous QCD sum rule determinations, dating back many years
\cite{GA0}, were the subject of some controversy. We find it possible to
reproduce the experimental value of $g_{A}$ at $T=0$, which then
serves to normalize the finite temperature results. 
Finally, as a byproduct, we shall use this result to determine the
behaviour at finite temperature of the $SU(2)_{L} \times SU(2)_{R}$
Goldberger-Treiman relation
\begin{equation}
  \frac{f_\pi(T) g_{\pi NN}(T)} {M_N(T) g_A(T)} = 1 + \Delta_{\pi}(T) \;\;.
\end{equation}
In this relation, $f_{\pi}(T)$ is known up to $T=T_{c}$ \cite{BAR1}, where
it vanishes, $g_{\pi NN} (T)$ behaves qualitatively similarly \cite{GPINNT},
and $M_N(T)$ is essentially constant up to $T=T_{c}$ \cite{MNT1}-\cite{MNT2}.
The question is then, how big are the thermal corrections to this relation,
$\Delta_{\pi}(T)$ (normalized to $\Delta_{\pi}(0)=0$).
An equally important chiral-symmetry relation, the Gell-Mann,
Oakes and Renner relation (GMOR), has recently been investigated in the
framework of thermal chiral perturbation theory \cite{TOUBLAN} and
QCD sum rules \cite{GMORT}. There is excellent numerical agreement between
both results, indicating that temperature corrections to the GMOR
relation are rather small. It should be kept in mind that a comparison
between thermal QCD sum rules results and those from effective theories at
finite temperature, e.g. sigma model, chiral perturbation theory, etc., must
necessarily be done numerically. The fields involved in
the former technique are those of the quarks and gluons, while those of
the latter framework are purely hadronic. As a result, expansions in powers
of the temperature do not necessarily need to match order by order because
the coefficients in these expansions will involve different types of
parameters. However, numerical results from both techniques should agree,
at least within the range of validity of the low temperature expansion in
effective theories (QCD sum rules are in principle valid across the whole
range of temperatures). This is precisely what happens with the two
analyses of the GMOR relation mentioned above.\\

We begin by considering the three-point function
\begin{equation}
  \Pi_\mu(p,p',q) = i^2 \dintst \langle 0 \left| T\left(\eta_p(x)
      A_\mu(y) \overline{\eta}_n(0) \right) \right|0 \rangle
  e^{i(p'x-qy)} \; ,
\end{equation}
where the charged axial vector current is given by:
$A_\mu(x) = \overline{u}(x) \gamma_\mu \gamma_5 d(x)$,
while the interpolating currents of the proton and neutron are chosen as
\cite{RRY}
\begin{eqnarray}
    \eta_p(x) &=& \epsilon_{abc} \left[ u^a(x) C \gamma_\mu u^b(x) \right]
  \gamma^\mu \gamma_5 d^c \; , \nonumber \\
   \eta_n(x) &=& -\epsilon_{abc} \left[ d^a(x) C \gamma_\mu d^b(x) \right]
  \gamma^\mu \gamma_5 u^c. 
\end{eqnarray}
The axial-vector coupling of the nucleon, $g_{A}(q^{2})$, is defined through
\begin{equation}
  \langle N(p_2) | A_\mu(0) | N(p_1) \rangle  = 
  \overline{u}(p_2) \left[ \gamma_\mu \gamma_5 \,g_A (q^2) + q_\mu
    \gamma_5 \,h_A(q^2) \right] u(p_1) \;, 
\end{equation}
with $q_\mu = (p_2 - p_1)_\mu$.
The coupling of the  interpolating currents, Eq.(3), to the nucleon is   
\begin{equation}
  \langle0 | \eta(0) | N(p) \rangle  = \lambda_N u(p) \;.
\end{equation}
Inserting a complete set of intermediate nucleon states into Eq.(2), one
obtains the hadronic representation
\begin{equation}
  \Pi_\mu(p,p',q) = \frac{\lambda_N^2}{(p^2 - M_N^2)(p'^2 - M_N^2)}
  (\ppslash + M_N ) T_\mu (\pslash + M_N ) \;,
\end{equation}
where
\begin{equation}
  T_\mu = \left[\gamma_\mu \gamma_5 \,g_A(q^2) + q_\mu \gamma_5
    \,h_A(q^2) \right] \; ,
\end{equation}
and the following expansion holds
\begin{eqnarray}
  (\ppslash + M_N)T_\mu(\pslash + M_N) & = & g_A(q^{2}) \left[ - 2 \ppslash
    p_\mu \gamma_5 + \ppslash \pslash \gamma_\mu \gamma_5 + (\pslash +
    \ppslash) \gamma_\mu \gamma_5 M_N \right. \nonumber \\
    & & \mbox{} \left. - 2 p_\mu \gamma_5 M_N + M_N^2
    \gamma_\mu \gamma_5 \right] \nonumber \\
  && \mbox{} + h_A(q^{2}) \left[ -\ppslash \pslash + \qslash \, M_N + M_N^2
  \right] q_\mu \gamma_5.
\end{eqnarray}
Since we are only interested in $g_A$, we need to extract tensor
structures which are not multiplied by $h_A$; a suitable candidate being
the structure
\begin{equation}
  (\ppslash + \pslash) \gamma_\mu \gamma_5 \; .
\end{equation}
The relevant term of the imaginary part of the (hadronic) correlator is then
\begin{eqnarray}
  Im \; \Pi_{\mu}(s,s',q^2)|_\mathrm{HAD} &=& 
  -\lambda_N^2 g_A(q^{2}) M_N \pi^2
   \delta(s
  - M_N^2) \delta(s' - M_N^2) (\ppslash+\pslash) \gamma_\mu \gamma_5
  \nonumber \\ & &
  \mbox{} + \Theta(s - s_0) \Theta(s' - s'_0) \;Im \; 
  \Pi_{\mu}(s,s',q^2)|_\mathrm{QCD} \; ,
\end{eqnarray}
where $s=p^{2}$, $s'=p'^{2}$, and
we have added the hadronic continuum, modelled by perturbative QCD,
starting at thresholds $s=s_{0}$ and $s'=s'_{0}$. 
Considering the contribution to the correlator from perturbative QCD,
we obtain
\begin{eqnarray}
 \Pi_\mu(p,p',q)|_\mathrm{PQCD} & = & - 24 i^2 \difour \gamma_\alpha S_F(k_2)
  \gamma_\beta S_F(k_1 - q) \gamma_\mu \nonumber \\
  & & \ \ \ \ \ \ \ \ \times S_F(k_1) \gamma^\alpha S_F(p' - k_1 -
  k_2) \gamma^\beta \gamma_5.
\end{eqnarray}
Taking the imaginary part of this expression, and
evaluating the integrals, it turns out that there are no terms
proportional to the tensor structure of Eq.(9). 
Turning to the non-perturbative part, we find the quark condensate
contribution to the correlator to be
\begin{eqnarray}
  \Pi_{\mu}(p,p',q)|_\mathrm{QCD} &=& 2 i^3 \langle 
  \overline{d} d \rangle \left[ \ifour
    \gamma_\alpha S_F(k) \gamma_\beta \gamma_\mu S_F(q) \gamma^\alpha
    S_F(p' - k - q) \gamma^\beta \gamma_5 \right. \nonumber \\
  & & \ \ \ \ \ \ \  \left. \mbox{} - \ifour\gamma_\alpha \gamma_\beta
   S_F(k - q)
    \gamma_\mu S_F(k) \gamma^\alpha S_F(p' - k) \gamma^\beta \gamma_5
  \right] \nonumber \\ & & 
    \mbox{} + \langle \overline{u} u \rangle \left[\ifour \gamma_\alpha
     S_F(p' -
      k) \gamma_\beta S_F(k - q) \gamma_\mu S_F(k) \gamma^\alpha
      \gamma^\beta \gamma_g \right. \nonumber \\ & &
    \ \ \ \ \  \ \ \left. \mbox{} + \ifour \gamma_\alpha S_F(k)
      \gamma_\beta S_F(q) \gamma_\mu \gamma^\alpha S_F(p' - k)
      \gamma^\beta \gamma_5 \right].
\end{eqnarray}
Taking the imaginary part, and keeping
only terms proportional to the relevant tensor structure
Eq.(9), and  which are non-vanishing in the limit $q^2 \rightarrow
0$, we obtain, after assuming $\langle \bar{u} u \rangle \simeq
\langle \bar{d} d \rangle \equiv \langle \bar{q} q \rangle$,
\begin{equation}
  Im \; \Pi_{\mu}(p,p',q)|_\mathrm{QCD} =  \frac{1}{12 \pi} 
  \langle \overline{q} q \rangle
  \left( \pslash + \ppslash \right) \gamma_\mu \gamma_5.
\end{equation}
Next, using Cauchy's theorem, and assuming quark-hadron duality, the
lowest dimensional FESR for $g_A$ reads
\begin{equation}
\dintsof \;Im \; \Pi_{\mu}(s,s')_\mathrm{HAD} =
\dintsof \;Im \; \Pi_{\mu}(s,s')_\mathrm{QCD}\;. 
\end{equation}
From this FESR one then obtains the relation
\begin{equation}
  g_A = -\frac{s_0 s'_0}{12 \pi^3} \frac{\langle \overline{q} q
    \rangle}{\lambda_N^2 M_N}\;.\\
\end{equation}
At first sight, this result hardly looks like a prediction for $g_{A}$,
since $s_{0}$, $s'_{0}$, and $\lambda_{N}$ are a-priori unknown. However,
since the double dispersion in $p^2 = s$ and $p'\;^2 = s'$, used in
obtaining Eq.(15), refers to the nucleonic legs of the three-point
function, it is reasonable to set $s_0 = s'_0$. At the same time, a QCD FESR
analysis of the two-point function involving nucleonic currents
\cite{MNT1} yields the following relations
\begin{equation}
  \lambda_N^2 = \frac{s_0^3}{192 \pi^4} \; \; , \; \;
  \lambda_N^2 M_N = - \frac{\langle \overline{q} q\rangle}{8 \pi^2} s_0^2
  \; ,
\end{equation}
where, in principle, the numerical value of the asymptotic freedom threshold
$s_{0}$ does not have to be the same as that in Eq.(15). In fact,
if one were to assume them to be equal, then Eqs.(13) and
(14) would imply $g_{A} = 8/12\pi$, a value far too small.
Without any attempt at extracting a {\it precision} value of $g_{A}$, it
is rewarding, though, to find that the experimental value $g_A = 1.26$
can be reproduced in this framework if
$\lambda_N^2 \simeq 3.1 \times 10^{-4}\ GeV^6$, $s_0 = 3.7\ GeV^2$,
and the standard value $\langle \bar{q} q \rangle = -0.01$, are used in Eq.
(15). These values of $\lambda_N^2$ and $s_0$ are close enough
to those resulting from the two-point function channel, at least for the
purpose of the present work, which is to obtain the temperature
dependence of $g_{A}$ and its mean square radius, rather than a prediction
for $g_{A}(T=0)$.\\

The finite temperature corrections to $g_A$ are obtained by inserting
the thermal  Dolan and Jackiw \cite{DOLJAK} propagators, 
and allowing for the temperature variation of $\langle \bar{q}
q\rangle$, $\lambda_N$ and $s_0$. For $\langle \bar{q} q\rangle _{T}$
and $\lambda_N(T)$
we shall use the results of \cite{BAR1} and of \cite{MNT1}, respectively. 
The temperature dependence of $s_{0}$ was first obtained in \cite{DL1},
and later improved in \cite{BAR2}. It turns out that for a wide range
of temperatures not too close to $T_{c}$, say $T < 0.8 T_{c}$, the
following scaling relation holds to a good approximation
\begin{equation}
\frac{f_{\pi}^{2}(T)}{f_{\pi}^{2}(0)} \simeq
\frac{\langle \bar{q} q\rangle _{T}}{\langle \bar{q} q\rangle _{0}} \simeq
\frac{s_{0}(T)}{s_{0}(0)} \;.
\end{equation}
The appropriate contribution
to the thermally corrected QCD spectral function becomes
\begin{equation}
  Im \; \Pi_{\mu}(p,p',q) = \frac{\langle\overline{q} q\rangle}{48\pi}
    (\pslash + \ppslash)\gamma_\mu\gamma_5  \left[f(p,T) + f(p',T) \right]
    \; ,
\end{equation}
where
\begin{equation} 
  f(p,T) = \int_{-1}^1 dx \left[ 1 -  n_F\left(\frac{|p_0 - \abp
   x|}{2}\right) - n_F\left(\frac{|p_0 + \abp x|}{2}\right)\right],
\end{equation}
with $n_{F}(x) = (1+e^{x})^{-1}$, and $f(p',T)$ is similarly defined.
Finally, we obtain  the sum rule for $g_A$ at finite temperature:
\begin{equation}
  g_A(T) = - \frac{\langle\overline{q} q\rangle}{48\pi^3}
  \frac{1}{\lambda_N^2 M_N} \dintsoT \left[f(p,T) + f(p',T) \right].
\end{equation}
In order to  evaluate the integrals one needs to choose a specific frame, 
for example the (rest) frame $\vec{p} = 0$. In this case, the 
components of the four vectors 
$p$ and $p'$ may be expressed in terms of $s$, $s'$ and $q^2$. Other
choices of frames give  essentially the same results.
A numerical
evaluation of $g_A(T)$ is presented in Fig. 1.
As can be seen from this figure, $g_A$ is basically T-independent, and
it clearly
does not vanish as the critical temperature is approached. In this sense,
$g_A$ does not represent a signal for the deconfinement phase transition.
We turn now  to the mean square radius $\langle r^2_A \rangle_T$
associated with $g_{A}$, and defined as
\begin{equation}
  \langle r^2_A \rangle_T = 6 \frac{\partial}{\partial q^2} \ln g_A(q^2,
  T)|_{q^2 = 0} \; .
\end{equation}
This radius is non-zero at finite temperature due to the $q^{2}$-dependence
of the arguments of the thermal Fermi factors. After evaluating the
logarithmic derivative of Eq.(20) one obtains
\begin{eqnarray}
  \langle r^2_A \rangle_T &=& \left\{\dintsof \left[f(p,T) +
      f(p',T)\right]\right\}^{-1} \dintsoT \int_{-1}^1 dx 
   \nonumber \\ 
  & &\times \frac{6}{2 T \sqrt{s}} \left( 1 + x \frac{p'_0}{\left|\vec{p}\;'
   \right|} \right)
  e^{\frac{p'_0 + \left|\vec{p}\;' \right| x}{2 T}} \left[ n_F\left(\frac{p'_0 
        + \left|\vec{p}\;'\right| x}{2} \right)\right]^2. 
\end{eqnarray}
This is plotted in fig.2, which shows that
the radius diverges as the critical temperature is approached. This kind
of behaviour has been obtained previously for other radii \cite{3PFT},
\cite{GPINNT}, and it may be interpreted as (analytic) evidence for quark 
deconfinement.\\

Finally, we can  use our result for $g_A$ at non-zero
temperature to evaluate the validity of the GTR, Eq.(1).
Results for the mass of the nucleon show that it has very little
variation with temperature, and so we shall assume that it is
constant \cite{MNT1}-\cite{MNT2}. Using the result of \cite{BAR1}
for $f_\pi$ at finite
temperature, together with our previous results for $g_{\pi NN}(T)$
\cite{GPINNT},
and our current result for $g_A(T)$, we can determine the thermal 
correction to the
GTR, $\Delta_{\pi}(T)$ defined in Eq.(1). In fig.3
we present a plot of $1 + \Delta_{\pi}(T)$ against $T/T_c$, which
indicates that the GTR is
approximately correct until about $T \simeq 0.9 \;T_c$, where
it breaks down.\\

\begin{large}
{\bf Figure Captions}\\
\end{large}

\noindent
Figure 1. The coupling $g_{A}(T)$, from Eq. (20), as a function
of $T/T_{c}$.\\

\noindent
Figure 2. The temperature dependence of the mean square radius, Eq.(22).\\

\noindent
Figure 3. Deviation from the Goldberger-Treiman relation, Eq.(1), as
          a function of of $T/T_{c}$.\\
\newpage

\begin{figure}[htbp]
  \begin{center}
    \vspace{30mm}
\setlength{\unitlength}{0.1bp}
\begin{picture}(3600,2160)(0,0)
\put(2050,150){\makebox(0,0){$T/T_c$}}
\put(100,1230){%
\makebox(0,0)[b]{\shortstack{$g_A(T)$}}%
}
\put(3407,300){\makebox(0,0){$1$}}
\put(3121,300){\makebox(0,0){$0.9$}}
\put(2836,300){\makebox(0,0){$0.8$}}
\put(2550,300){\makebox(0,0){$0.7$}}
\put(2264,300){\makebox(0,0){$0.6$}}
\put(1979,300){\makebox(0,0){$0.5$}}
\put(1693,300){\makebox(0,0){$0.4$}}
\put(1407,300){\makebox(0,0){$0.3$}}
\put(1121,300){\makebox(0,0){$0.2$}}
\put(836,300){\makebox(0,0){$0.1$}}
\put(550,300){\makebox(0,0){$0$}}
\put(500,2060){\makebox(0,0)[r]{$3$}}
\put(500,1783){\makebox(0,0)[r]{$2.5$}}
\put(500,1507){\makebox(0,0)[r]{$2$}}
\put(500,1230){\makebox(0,0)[r]{$1.5$}}
\put(500,953){\makebox(0,0)[r]{$1$}}
\put(500,677){\makebox(0,0)[r]{$0.5$}}
\put(500,400){\makebox(0,0)[r]{$0$}}
\end{picture}
          \caption{}
    \label{fig:gAT}
  \end{center}
\end{figure}
\newpage
\begin{figure}[htbp]
  \begin{center}
    \vspace{30mm}
\setlength{\unitlength}{0.1bp}
\begin{picture}(3600,2160)(0,0)
\put(2075,150){\makebox(0,0){$T/T_c$}}
\put(100,1230){%
\makebox(0,0)[b]{\shortstack{$\langle r^2_A \rangle_T$}}%
}
\put(3410,300){\makebox(0,0){$1$}}
\put(3129,300){\makebox(0,0){$0.9$}}
\put(2848,300){\makebox(0,0){$0.8$}}
\put(2567,300){\makebox(0,0){$0.7$}}
\put(2286,300){\makebox(0,0){$0.6$}}
\put(2005,300){\makebox(0,0){$0.5$}}
\put(1724,300){\makebox(0,0){$0.4$}}
\put(1443,300){\makebox(0,0){$0.3$}}
\put(1162,300){\makebox(0,0){$0.2$}}
\put(881,300){\makebox(0,0){$0.1$}}
\put(600,300){\makebox(0,0){$0$}}
\put(550,2060){\makebox(0,0)[r]{$2$}}
\put(550,1728){\makebox(0,0)[r]{$1.5$}}
\put(550,1396){\makebox(0,0)[r]{$1$}}
\put(550,1064){\makebox(0,0)[r]{$0.5$}}
\put(550,732){\makebox(0,0)[r]{$0$}}
\put(550,400){\makebox(0,0)[r]{$-0.5$}}
\end{picture}
    \caption{}
    \label{fig:radga}
  \end{center}
\end{figure}
\newpage
\begin{figure}[htbp]
  \begin{center}
    \vspace{30mm}
\setlength{\unitlength}{0.1bp}
\begin{picture}(3600,2160)(0,0)
\put(2050,150){\makebox(0,0){$T/T_c$}}
\put(100,1230){%
\makebox(0,0)[b]{\shortstack{$1 + \Delta_\pi(T)$}}%
}
\put(3407,300){\makebox(0,0){$1$}}
\put(3121,300){\makebox(0,0){$0.9$}}
\put(2836,300){\makebox(0,0){$0.8$}}
\put(2550,300){\makebox(0,0){$0.7$}}
\put(2264,300){\makebox(0,0){$0.6$}}
\put(1979,300){\makebox(0,0){$0.5$}}
\put(1693,300){\makebox(0,0){$0.4$}}
\put(1407,300){\makebox(0,0){$0.3$}}
\put(1121,300){\makebox(0,0){$0.2$}}
\put(836,300){\makebox(0,0){$0.1$}}
\put(550,300){\makebox(0,0){$0$}}
\put(500,1981){\makebox(0,0)[r]{$1$}}
\put(500,1665){\makebox(0,0)[r]{$0.8$}}
\put(500,1349){\makebox(0,0)[r]{$0.6$}}
\put(500,1032){\makebox(0,0)[r]{$0.4$}}
\put(500,716){\makebox(0,0)[r]{$0.2$}}
\put(500,400){\makebox(0,0)[r]{$0$}}
\end{picture}
    \caption{}
    \label{fig:gtrft}
  \end{center}
\end{figure}

\end{document}